\PassOptionsToPackage{unicode}{hyperref}
\PassOptionsToPackage{hyphens}{url}
\documentclass[
  manuscript,
  screen,
  authorversion,
  nonacm]{acmart}
\usepackage{amsmath}
\usepackage{iftex}
\setcopyright{cc}
\ifPDFTeX
  \usepackage[T1]{fontenc}
  \usepackage[utf8]{inputenc}
  \usepackage{textcomp} 
\else 
  \usepackage{unicode-math} 
  \defaultfontfeatures{Scale=MatchLowercase}
  \defaultfontfeatures[\rmfamily]{Ligatures=TeX,Scale=1}
\fi
\usepackage{lmodern}
\ifPDFTeX\else
\fi
\IfFileExists{upquote.sty}{\usepackage{upquote}}{}
\IfFileExists{microtype.sty}{
  \usepackage[]{microtype}
  \UseMicrotypeSet[protrusion]{basicmath} 
}{}
\makeatletter
\@ifundefined{KOMAClassName}{
  \IfFileExists{parskip.sty}{%
    \usepackage{parskip}
  }{
    \setlength{\parindent}{0pt}
    \setlength{\parskip}{6pt plus 2pt minus 1pt}}
}{
  \KOMAoptions{parskip=half}}
\makeatother
\usepackage{xcolor}
\usepackage{graphicx}
\makeatletter
\def\maxwidth{\ifdim\Gin@nat@width>\linewidth\linewidth\else\Gin@nat@width\fi}
\def\maxheight{\ifdim\Gin@nat@height>\textheight\textheight\else\Gin@nat@height\fi}
\makeatother
\setkeys{Gin}{width=\maxwidth,height=\maxheight,keepaspectratio}
\makeatletter
\def\fps@figure{htbp}
\makeatother
\providecommand{\tightlist}{%
  \setlength{\itemsep}{0pt}\setlength{\parskip}{0pt}}
\NewDocumentCommand\citeproctext{}{}

\makeatletter
 \let\@cite@ofmt\@firstofone
 \def\@biblabel#1{}
 \def\@cite#1#2{{#1\if@tempswa , #2\fi}}
\makeatother
\newlength{\cslhangindent}
\newlength{\csllabelwidth}
\newenvironment{CSLReferences}[2] 
 {\begin{list}{}{%
  \setlength{\itemindent}{0pt}
  \setlength{\leftmargin}{0pt}
  \setlength{\parsep}{0pt}
  \ifodd #1
   \setlength{\leftmargin}{\cslhangindent}
   \setlength{\itemindent}{-1\cslhangindent}
  \fi
  \setlength{\itemsep}{#2\baselineskip}}}
 {\end{list}}
\usepackage{calc}

\newcommand{\CSLLeftMargin}[1]{\parbox[t]{\csllabelwidth}{\strut#1\strut}}
\newcommand{\CSLRightInline}[1]{\parbox[t]{\linewidth - \csllabelwidth}{\strut#1\strut}}

\author{A. Samuel Pottinger}
\orcid{0000-0002-0458-4985}
\email{sam.pottinger@berkeley.edu}
\affiliation{
  \institution{Schmidt Center for Data Science and Environment, University of California}
  \city{Berkeley}
  \state{CA}
  \country{USA}
}

\author{Nivedita Biyani}
\affiliation{
  \institution{Benioff Ocean Science Laboratory, Marine Science Institute, University of California}
  \city{Santa Barbara}
  \state{CA}
  \country{USA}
}

\author{Roland Geyer}
\orcid{0000-0001-7430-1058}
\affiliation{
  \institution{Bren School of Environmental Science and Management, University of California}
  \city{Santa Barbara}
  \state{CA}
  \country{USA}
}

\author{Douglas J McCauley}
\orcid{0000-0002-8100-653X}
\affiliation{
  \institution{Schmidt Center for Data Science and Environment, UCB}
  \city{Berkeley}
  \state{CA}
  \country{USA}
}
\affiliation{
  \institution{Benioff Ocean Science Laboratory, Marine Science Institute, UCSB}
  \city{Santa Barbara}
  \state{CA}
  \country{USA}
}
\affiliation{
  \institution{Ecology, Evolution, and Marine Biology Department, UCSB}
  \city{Santa Barbara}
  \state{CA}
  \country{USA}
}
\affiliation{
  \institution{Department of Environmental Science, Policy and Management, UCB}
  \city{Berkeley}
  \state{CA}
  \country{USA}
}

\author{Magali de Bruyn}
\orcid{0009-0000-1609-6244}
\affiliation{
  \institution{Schmidt Center for Data Science and Environment, University of California}
  \city{Berkeley}
  \state{CA}
  \country{USA}
}

\author{Molly R Morse}
\orcid{0000-0002-6961-6503}
\affiliation{
  \institution{Benioff Ocean Science Laboratory, Marine Science Institute, University of California}
  \city{Santa Barbara}
  \state{CA}
  \country{USA}
}

\author{Neil Nathan}
\orcid{0000-0003-3945-8484}
\affiliation{
  \institution{Benioff Ocean Science Laboratory, Marine Science Institute, University of California}
  \city{Santa Barbara}
  \state{CA}
  \country{USA}
}

\author{Kevin Koy}
\orcid{0000-0002-8479-4404}
\affiliation{
  \institution{Schmidt Center for Data Science and Environment, University of California}
  \city{Berkeley}
  \state{CA}
  \country{USA}
}

\author{Ciera Martinez}
\orcid{0000-0003-4296-998X}
\affiliation{
  \institution{Schmidt Center for Data Science and Environment, University of California}
  \city{Berkeley}
  \state{CA}
  \country{USA}
}

\begin{abstract}
\textbf{Introduction}: This multi-disciplinary case study details how an interactive decision support tool leverages game design to inform an international plastic pollution treaty.

\textbf{Design}: Seeking to make our scientific findings more usable within the policy process, our interactive software supports manipulation of a mathematical model using techniques borrowed from games. These "ludic" approaches aim to enable user agency to find custom policy solutions, invite deep engagement with scientific results, serve audiences of diverse expertise, and accelerate scientific process to keep pace with intergovernmental negotiations.

\textbf{Implementation}: Built in JavaScript and D3 with user-modifiable logic via an ANTLR domain specific language, this browser-based application offers adaptability and explorability for our machine learning results with privacy preserving architecture and offline capability.

\textbf{Demonstration}: Policymakers and the supporting community engaged with this public simulation tool across multiple treaty-related events, investigating plastic waste outcomes under diverse and sometimes unexpected policy scenarios.

\textbf{Conclusion}: Contextualizing our open source software within a broader lineage of digital media research, we reflect on this interactive modeling platform, considering how game design approaches may help facilitate collaboration at the science / policy nexus.

\textbf{Materials}: Available on the public Internet, we host this browser-based decision support tool at global-plastics-tool.org, work also archived at zenodo.org/records/12615011 in a Docker container.

\end{abstract}
\ifLuaTeX
  \usepackage{selnolig}  
\fi
\usepackage{bookmark}
\IfFileExists{xurl.sty}{\usepackage{xurl}}{} 
\hypersetup{
  pdftitle={Using Game Design to Inform a Plastics Treaty: Fostering Collaboration between Science, Machine Learning, and Policymaking},
  pdfauthor={A. Samuel PottingerNivedita BiyaniRoland GeyerDouglas J
McCauleyMagali de BruynMolly R MorseNeil NathanKevin KoyCiera Martinez},
  hidelinks,
  pdfcreator={LaTeX via pandoc}}

\title{Using Game Design to Inform a Plastics Treaty: Fostering
Collaboration between Science, Machine Learning, and Policymaking}
\author{}
\date{2025-11-04}

\begin{document}
\maketitle

\section{Introduction}\label{introduction}

In his influential ``Humane Representation of Thought'' presentation,
Bret Victor {[}79{]} explores how interactivity enables new ways of
reasoning to unlock thoughts we ``couldn't think'' before. This paper
explores one such application of these ideas: the open source
\href{https://global-plastics-tool.org}{global-plastics-tool.org}.
Constructed by this article's cross-disciplinary research team as a
complement to our traditional academic publication {[}57{]}, this
decision support tool seeks to facilitate iterative policy exploration
beyond the capabilities of a static paper. Specifically, we look to game
design in crafting this interactivity, attempting to improve usability
of our scientific insights for negotiators and stakeholders engaging in
the development of the United Nations' ``International Legally Binding
Instrument on Plastic Pollution, Including in the Marine Environment''
{[}73{]}.

\subsection{Background}\label{background}

Our tool and this analysis participate in a broader dialogue in the
literature regarding interactive media to facilitate reasoning in
complex problems. This prior work may inform the construction of tools
which support policy decisions like designing this
``once-in-a-lifetime'' global treaty {[}42{]}.

\subsubsection{Explorable explanations}\label{explorable-explanations}

Building on work reaching back to 1960s PLATO {[}9{]}, Victor draws
attention to the possibilities of ``explorable explanations'' for
``active'' reading. He observes that, while static media cannot invite
response and leaves ``the reader's line of thought'' both ``internal and
invisible'' {[}77{]}, new interactive technologies allow readers to
``as{[}k{]} questions, conside{[}r{]} alternatives, questio{[}n{]}
assumptions, and even questio{[}n{]} the trustworthiness of the
author.'' These design ideas have been applied in education,
engineering, and industry to improve user agency and bolster the impact
of modeling efforts {[}11, 22, 24, 32{]}.

\subsubsection{Scientific
communications}\label{scientific-communications}

However, years after the formulation of these concepts for interactivity
and knowledge building, Dragicevic et al. {[}27{]} write that Victor's
ideas still remain uncommon with ``very few concrete examples'' in
academic scientific communication. Continuing to speak in unresponsive
media of papers and slides, the outputs of research still often cannot
listen to stakeholders' perspectives or react to their ideas {[}79{]},
missing out on ``the deep understanding that comes from dialogue and
exploration'' {[}77{]}.

\subsubsection{Related work}\label{related-work}

Among the relatively few projects trying to realize this power of
interactivity in scientific communication {[}27{]}, we observe some
recent examples in environment and sustainability:

\begin{itemize}
\tightlist
\item
  Driving the Future: Simulation of automotive markets under different
  incentive, tax, and policy scenarios {[}40{]}.
\item
  En-ROADS: Interactive simulation for policy impact on greenhouse gas
  emissions and global warming {[}36, 60{]}.
\item
  Protect High Seas: Tool to identify potential marine protected area
  candidates using configurable conservation plans {[}80{]}.
\item
  MSP Challenge: A multi-user environment for collaborative marine
  spatial planning with simulation of policy effects {[}3{]}.
\end{itemize}

Enabling co-design of science-informed solutions, all of these
exceptional digital environments turn static information into engaging
software which allow users to bi-directionally converse with data.

\subsection{Contribution}\label{contribution}

We build upon prior work by offering an additional documented example of
these ideas within the context of international policymaking for plastic
pollution, distilling design vocabulary for decision support tools with
a focus on four key hurdles in the policy / science gap {[}71{]}.

\subsubsection{Challenge 1: User agency}\label{challenge-1-user-agency}

First, Chris Tyler of the United Kingdom's Parliamentary Office of
Science and Technology warns that scientists ``often consider themselves
as the `experts' who engage with policy makers'' but they should not
assume ``that {[}they{]} are the only expert{[}s{]} in the room''
{[}71{]}. Jan Nolin's investigation of the process of formulating the
Kyoto Proposal similarly highlights the benefits of bidirectional
engagement in drafting legislation, underscoring the necessity of
bringing expertise from science and policy into dialogue in order to
craft effective solutions {[}49{]}. With this need for mutual
collaboration in mind, consider that the typical medium of scientific
communication in the form of a static academic paper may limit
policymaker agency and co-creation opportunities. By ``display{[}ing{]}
the author's argument, and nothing else,'' static media only support
one-way discussion, possibly hindering the reader's agency to evaluate
questions or alternative ideas {[}77{]}.

\subsubsection{Challenge 2:
Approachability}\label{challenge-2-approachability}

Simulation of policy impacts often requires exploring effects across
many dimensions {[}36, 60{]}. However, this can be taxing as
visualization of highly multi-variate data often encounter ``perceptual
and cognitive limits'' {[}29{]}. This presents a design challenge in
maintaining the required complexity without tools feeling complicated to
the point of becoming inaccessible {[}7{]}.

\subsubsection{Challenge 3: Reading
levels}\label{challenge-3-reading-levels}

Recommendations to the scientific community on policy engagement often
emphasize the need for brief summaries in plain language {[}33{]}.
However, information designers conversely warn that traditional media
like slides cause oversimplification and result in poor decision making
{[}70{]}. Policy decision support tools may step into this tension:
attempting to serve audiences of diverse prior experience to support
both high level reading like the overviews recommended to scientists in
decision-maker engagement while simultaneously inviting deeper reading
when demanded by the policymaking process.

\subsubsection{Challenge 4: Speed}\label{challenge-4-speed}

Finally, Nolin observes that ``scientific work is usually slow while the
pace of policy is erratic'' {[}49{]}. To understand why, consider that
negotiation theory believes collaboration happens across multiple
``phases'' of information exchange {[}4{]} so the needs and questions of
stakeholders naturally adjust as perspective evolves. In other words,
``scientific input to be effective must thus be timed for readiness when
a new phase of policy shifts in'' {[}49{]}. In this context, both Nolin
and Tyler note that policy ``decisions usually need to be made pretty
quickly'' with strong constraints on timeline for additional research
{[}71{]}. Again, this encounters an inherent limitation of static media
on audience agency where readers can ``form questions, but can't answer
them'' {[}77{]}. To this end, Victor considers how digital alternatives
may respond to inquiry in the moment itself, addressing concerns
otherwise only ``answered by a phone call or heavy research'' {[}77{]}.
Altogether, policy-focused scientific work may need computation's
ability to ``react'' in order to keep pace with shifting dialogue
{[}78{]}.

\bigskip 

\section{Design}\label{design}

To solve for these four challenges in our treaty decision support tool,
we observe that interactive data visualization has previously looked to
game design for inspiration. For example, Elijah Meeks applies game
world structure concepts {[}45{]} while Pottinger et al.~explore ludic
techniques for tutorialization and agency {[}56, 59{]}. We continue that
work in using game design to structure the application of different
interaction patterns.

\subsection{Design for user agency}\label{design-for-user-agency}

Though interaction techniques identified by Victor seek to foster model
exploration {[}27, 77{]}, simply exposing our many parameters means that
science tools like this often require ``sophistication to navigate''
{[}59{]}. If this complexity feels overwhelming, it may inhibit the
agency that interactivity otherwise provides. However, video games must
also similarly introduce the concepts of a complex novel system. Indeed,
prior interactive scientific visualizations feature ``Hayashida-style''
tutorials {[}59{]}, leveraging a powerful 4-step technique common to
Nintendo's work {[}19, 50{]}:

\begin{enumerate}
\def\labelenumi{\arabic{enumi}.}
\tightlist
\item
  \textbf{Introduction}: Show the mechanic to the user.
\item
  \textbf{Development}: Give players a simplified task to try the
  action.
\item
  \textbf{Twist}: Highlight something unexpected about the skill.
\item
  \textbf{Conclusion}: Challenge users to demonstrate mastery of the
  lesson.
\end{enumerate}

Even so, tools like those from Pyafscgap using this structure may lead
through written tutorials in prescribed example analyses {[}59{]}. This
explicit guiding of the user may detract from a sense of freedom or make
the tool appear non-objective, asking if simulations can orient readers
without ``lecturing at'' them. This ``ludonarrative dissonance'' problem
also appears in games which, as detailed by Clint Hocking, arises when a
mechanic used by a player feels out of place with the experience's
intended message or narrative {[}34, 38{]}. Specifically, in our tool
for the plastic treaty, a written introductory tutorial may:

\begin{itemize}
\tightlist
\item
  Treat users like novices even as the tool invites them to bring their
  expertise to designing a solution.
\item
  Bias self-directed exploration by showing example insights even though
  the tool tries to present neutrality.
\end{itemize}

Believing users should discover solutions instead of having opinions
prescribed, we therefore look to Nintendo's 2017 ``The Legend of Zelda:
Breath of the Wild'' which offers a possible ``open world'' remedy. It
starts journeys in beginning areas with reduced complexity not
identified as a tutorial but which still serve the same purpose
{[}63{]}.

\begin{figure}
\centering
\includegraphics[width=0.83\textwidth,height=\textheight]{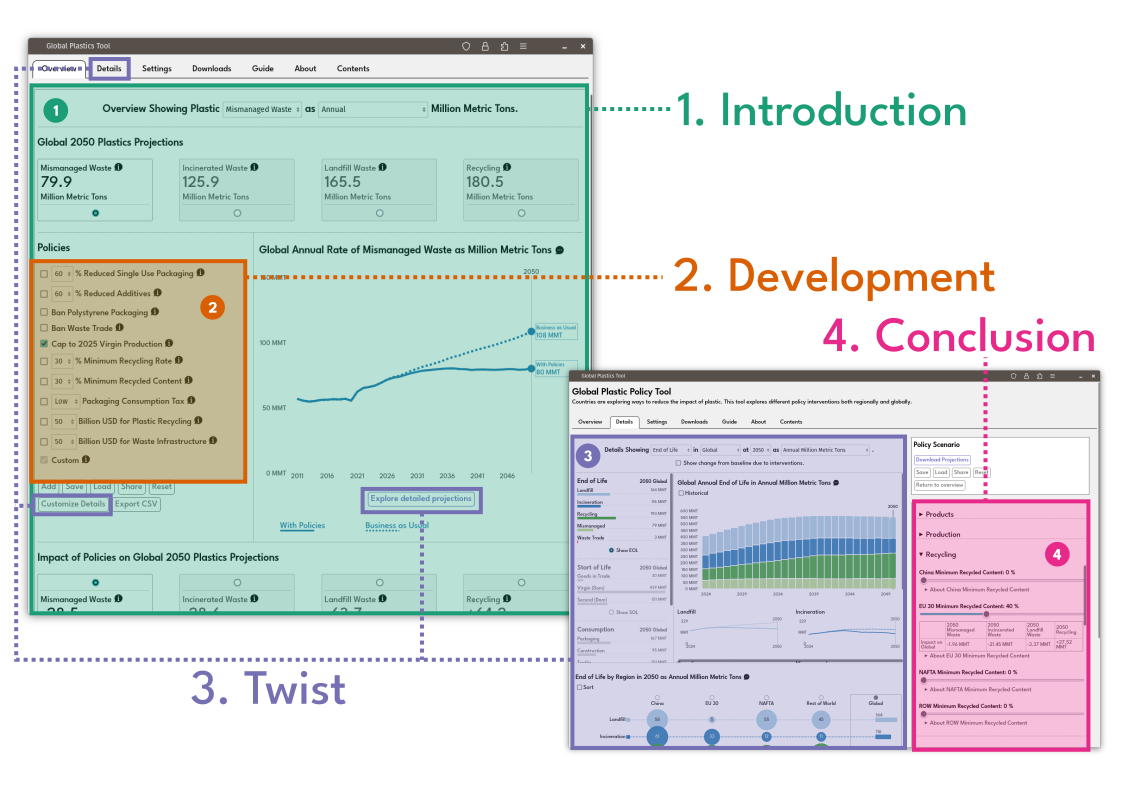}
\caption{Visualization structure embeds a Hayashida sequence. The
initial business as usual analysis without interaction provides an
introduction, the high level policy controls offer development, the
details tab adds additional dimensionality for the twist, and the
sliders refine initial policy selections for conclusion.}
\end{figure}

With Zelda in mind, this interactive tool still follows a Hayashida
sequence to foster tool familiarity required for user agency. However,
it uses visual hierarchy instead of narrative to construct a ``starting
region'' to introduce mechanics which blossom in the ``full world'' in
the details tab, making the tutorial ``invisible'' {[}15{]}.

\subsection{Design for multi-variate
evaluation}\label{design-for-multi-variate-evaluation}

While this introductory sequence considers how first impressions with
the tool may avoid a feeling of complexity, evaluating effects of
user-designed policy packages can span up to 90 views in the details
tab. To balance rich functionality with simplicity, our treaty support
software's design draws inspiration from video games' separation of
play:

\begin{itemize}
\tightlist
\item
  Games with \textbf{connected stage designs} like in Rare's
  ``Banjo-Tooie'' allow actions in one area to impact another even as it
  still separates content into discrete levels {[}20{]}. This creates a
  deep world in the ``secondary loop'' while building a ``primary
  gameplay loop'' {[}31{]} that naturally narrows the moment to moment
  available subset of that complexity.
\item
  Designer James Portnow observes that \textbf{open world games}
  advertise content through ``landmarks'' which orient players without
  explicit instruction {[}54{]}. As analyzed by Mark Brown, Zelda
  achieves this by peeking tall secondary loop landmarks over the
  world's hilly terrain which hide primary loop landmarks in distant
  areas {[}16{]}.
\end{itemize}

The tool operates in a way similar to a valley seen in Zelda. Though
users can only take a limited number of actions at any time related to
their current view in the primary loop, the tool keeps secondary loop
landmarks (like other lifecycle stages or regions) visible ``in the
distance'' through small previews opening into those other spaces. These
sight-lines try to encourage exploration and pull users ``over the
hill'' into a new valley. Furthermore, like Banjo-Tooie, changes in one
area reflect in others.

\begin{figure}
\centering
\includegraphics[width=0.83\textwidth,height=\textheight]{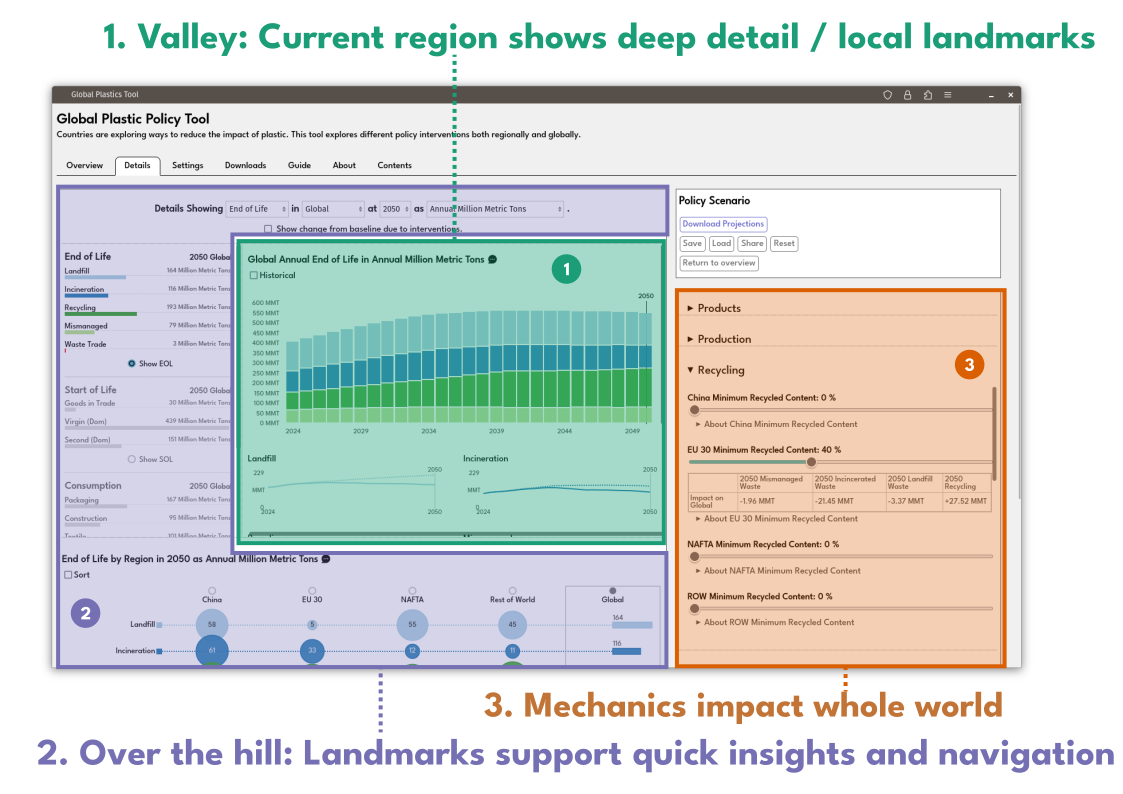}
\caption{The details tab uses ``valleys'' to structure primary and
secondary loops.}
\end{figure}

\subsection{Design for levels of
reading}\label{design-for-levels-of-reading}

Even if hidden tutorials and manipulated landmark sightlines introduce
the visualization tool's mechanics and orient users in a complex
multi-dimensional space, our software still serves various audiences
with different objectives and background familiarity. As Victor writes,
explorable explanations should consider how to empower users to ``read
at multiple levels'' of depth {[}77{]}. In other words, those
knowledgeable of plastics concepts or only seeking high level takeaways
may engage with the tool superficially while others may ``read closely''
to learn unfamiliar terminology or consider expert details.

To build these multiple simultaneous levels of engagement in our
interactive plastic model, producer Rob Nelson observes that a unique
``ability that games have and other mediums don't {[}is{]} optional,
additional narrative that you can choose to engage with or skip over''
{[}69{]}. In games:

\begin{itemize}
\tightlist
\item
  \textbf{Environmental storytelling} unobtrusively layers content into
  ``setting'' like graffiti or signage that the player encounters
  ``along their journey'' if they are paying close attention but can
  optionally be ignored, allowing the user to naturally choose their
  preferred depth of engagement {[}18, 23{]}.
\item
  \textbf{Rumors} which are subtle mentions of ``side-quests'' afford
  unobtrusive opt-in: invitations naturally missed by users who do not
  want that content or are busy rushing to another objective {[}17{]}.
\end{itemize}

These designs aim to support a deep path for those moving slowly and
another streamlined route for those moving quickly in which optional
elements blend into the background. Similarly, our interface prepares a
two track path that, while using Victor's interaction patterns {[}77{]},
leverages structural ideas from gaming to determine where to place those
elements like tooltips or drawers. Low within the visual hierarchy like
environmental storytelling, this architecture aims to allow those in a
rush to skip these ``rumors'' even as they still direct close readers
{[}17{]}.

\begin{figure}
\centering
\includegraphics[width=0.8\textwidth,height=\textheight]{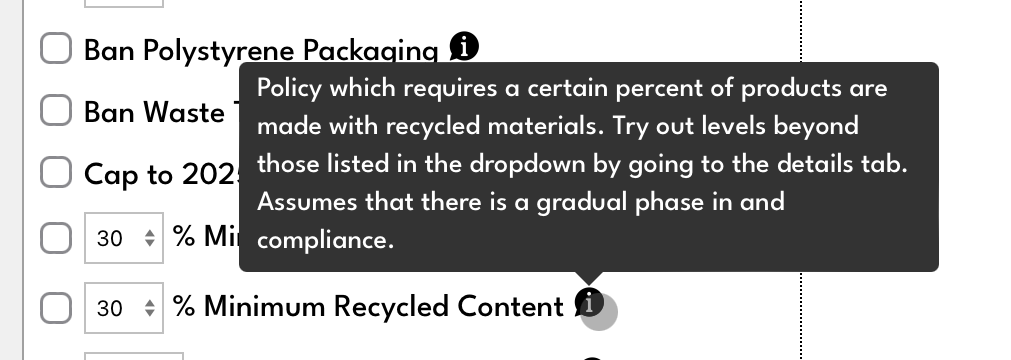}
\caption{A tooltip describes a technical phrase and beacons deeper.}
\end{figure}

\subsection{Design for speed}\label{design-for-speed}

A tool frequently used by games, this software turns to a domain
specific language (DSL) to accelerate defining policy logic {[}66,
82{]}. As informed by prior related projects {[}28, 41, 53, 58, 67{]},
our programming language includes task-specific keywords and operators
for common plastic simulation actions like calculating product lifecycle
distributions, proportionally distributing effects as part of
propagation, and supporting Monte Carlo. Relative to earlier prototypes
written in the R and Python programming languages, we sought the
following with this DSL:

\begin{itemize}
\tightlist
\item
  Improved readability of code through integration of domain specific
  keywords {[}37{]}.
\item
  Reduced length of scripts by removing ``boilerplate'' code {[}51{]}.
\item
  Shorter development cycles, making it easier to rapidly implement new
  policies to keep pace with policy negotiation {[}49{]}.
\end{itemize}

\bigskip 

\section{Deployment}\label{deployment}

Along with other smaller libraries mentioned within our
\href{https://github.com/SchmidtDSE/plastics-prototype}{project
documentation}, we implemented our tool using D3 {[}12{]} and ANTLR
{[}53{]}. This runs within the web browser using JavaScript with an
optional Progressive Web Application. We targeted our release to policy
makers, news media, and other organizations related to negotiating for
the United Nation's ``International Legally Binding Instrument on
Plastic Pollution, Including in the Marine Environment'' {[}76{]}.

In addition to sharing directly with government delegations as well as
other researchers and industry, our software was featured in either
official or unofficial ``side events'' in the Third and Fourth
Intergovernmental Negotiation Committee meetings for the treaty {[}72,
75{]} as well as at the Sixth Session of the United Nations
Environmental Assembly {[}74{]}. Furthermore, the tool was shared in
webinars and news articles {[}30, 62{]}. Finally, in addition to these
modalities (workshop, kiosk, bilateral, and presentation), web traffic
saw just over 13,000 daily page views\footnote{As described in our
  privacy considerations, these numbers do not filter by user agent
  string or de-duplicate by IP address.} during November 2023 largely
coinciding with news coverage and attention on social media {[}6, 30,
62{]}. In these engagements, individuals likely leveraged the tool
without us present. The software remains accessible at
\href{https://global-plastics-tool.org}{global-plastics-tool.org} as the
treaty process is expected to continue with a still unclear end date
{[}35{]}.

As expected by Nolin {[}49{]}, the tool's meetings and workshops
happened on the sidelines of an unpredictable fast-changing negotiation
{[}83{]}. As predicted by Victor {[}77{]}, we found the software helpful
in facilitating conversations at the science / policy nexus as we
navigated cycles of question and answer. Additionally, we found
interactivity helpful with on-the-spot exploring ideas and policy
configurations we did not originally anticipate in our scientific
modeling. With that in mind, we next further describe our experiences
and strategies for deploying this tool.

\subsection{Evaluative constraints}\label{evaluative-constraints}

As negotiation theory focuses on the importance of information exchanges
{[}4{]}, consider that evaluative approaches may obtain data about
insights within the context of the treaty {[}48{]}. For example,
Multi-dimensional In-depth Long-term Case-studies (MILC) would involve
the disclosure of not just experiences of the tool but also example
findings from the data {[}64{]}. These perspectives may become crucial
to negotiation.

Therefore, even if taking steps to ensure data safety, consideration of
the geopolitical context leads us to forgo collection of information
about individuals and specific treaty insights or actions within our
tool {[}42{]}. Respecting this practical need for elevated privacy both
enables this interactive science project to participate in this
multi-national process and offers a public scientific resource to the
community addressing a pressing global concern {[}84{]}, answering the
call to go beyond the paper and ``blend research and engagement'' to
``respond rapidly to real-world events'' {[}68{]}. This pragmatic stance
prevents this case study from providing findings from controlled
experiments or diary studies. Even so, we document our strategies as
facilitators in workshops and other collaborative meetings in hopes of
furthering the utility of our tool {[}26{]}.

\subsection{Development}\label{development}

During our engagements, our team developed strategies to keep pace with
the plastics policymaking process.

\subsubsection{Policy selection}\label{policy-selection}

First, we found that the political process offered moments of formal
documentation which we used to guide our development activities {[}4{]}.
In particular, the ``zero draft'' put out by the United Nations {[}76{]}
offered a snapshot on the state of thinking for the plastics treaty
which guided prioritization, ensuring support was available (where
possible) for policies it listed and deciding what to emphasize within
our interface. Future documents could similarly help our software keep
pace with evolving negotiations or implementation discussions.

\subsubsection{Integrated development
environment}\label{integrated-development-environment}

Despite the benefit of those formal documents, we still required fast
implementation loops predicted by Nolin {[}49{]}. For example, treaty
negotiation developments demanded rapid changes in the tool to support
new interventions, sometimes within 24 hours. To keep pace with policy
dialogue, our integrated code editor allowed us to quickly prototype
these policies, repurposing the sliders intended for user exploration to
``debug'' our implementations. This operationalized our domain specific
language for accelerated iteration. Similar to how the game design
communities find benefits both from DSLs and a ``hot reload'' to change
programming in-situ {[}25, 82, 85{]}, this live development capability
served not just during demonstrations but development of the tool as
well. Also, thanks to the DSL's brevity and built-in task-specific
operations, the web application often served as the first development
environment for a new policy, offering the ability to see how new logic
interacts with the model outputs in real time without the delay of a
``compile'' step. The ``prototype'' drawer in the details tab embodies
this pattern. Altogether, responding to Nolin's timeliness requirement,
this approach offered us agility as scientists and facilitators in
keeping pace with policy conversation {[}49{]}.

\subsection{Facilitation}\label{facilitation}

In addition to refining our development, we also evolved our
facilitation practices during deployment.

\subsubsection{Permission to play}\label{permission-to-play}

We quickly adopted an approach similar to that recommended by En-ROADS
{[}39{]} where workshops or presentations using the software feature a
``facilitator'' who first introduces the motivations and science behind
the tool. Offering ``permission to play'' {[}81{]}, we used this
opportunity to provide an invitation for policy makers to engage with
the results through their own inquiry.

\subsubsection{Progression}\label{progression}

Some early presentations tried listing all tool features to accelerate
conversation. However, our facilitators ultimately streamlined this
introduction and put users in front of the overview tab without
proactively mentioning the details tab. They felt that naturally arising
questions already draw engagement to the other parts of the tool: the
simpler overview tab surveys a broader possibility space at a high level
before conversation moves on its own to more deeply considering specific
modeling or policy aspects in the details tab.

\subsubsection{Live coding}\label{live-coding}

The DSL achieved sufficient brevity such that the visualization tool
could display policy logic within the details tab and allowed us to
discuss and modify that code during live demonstrations. This in-situ
programming capability creates what Maggie Appleton calls a
``Programming Portal'' {[}5{]}. In exploring user ideas during tool
demonstrations, this allowed us to, as Victor writes, ``play with the
\ldots{} assumptions and analyses {[}to{]} see the consequences'' in the
moment {[}77{]}, affording an important flexibility during conversation.

\subsubsection{Integrated Monte Carlo}\label{integrated-monte-carlo}

Not part of the original launch, questions in scientifically expert
discussions regarding sensitivity and uncertainty prompted us to add a
Monte Carlo feature which can explore output distributions under
different assumptions {[}46{]}. This further takes advantage of our
domain specific language.

\subsubsection{Progressive web
application}\label{progressive-web-application}

Pragmatically, we found heavy network utilization at these events as
well as uncertain connectivity at ``side-line'' locations. To that end,
we built a Progressive Web Application which can be installed or cached
to run offline.

\bigskip 

\section{Discussion}\label{discussion}

We next discuss a vocabulary of design patterns and further detail
implementation decisions before considering limitations and
opportunities for future work.

\subsection{Overview of design
patterns}\label{overview-of-design-patterns}

We first start discussion by summarizing and naming the design patterns
considered. However, recognizing the limitations on our evaluative
methods due to our tool's use case, this study encourages future
research of these ideas in evaluative contexts not possible in this
project.

\subsubsection{Structural Hayashida
Tutorials}\label{structural-hayashida-tutorials}

Building on prior work in both data visualization {[}59{]} and game
design {[}50{]}, visual hierarchy embeds the four step Hayashida
sequence into the architecture of the tool, constructing an
``invisible'' tutorial {[}15{]}. This narrative approach attempts to
orient users while avoiding the ludonarrative dissonance that an
explicit guide may generate {[}34, 38{]}. This tool's two tab connected
layout attempts to embody this possible instructional pattern.

\subsubsection{Connected Valleys}\label{connected-valleys}

The combination of open world and connected level designs deploys small
``landmark'' visualizations which both attempt to orient the player
navigationally and invite deeper exploration of alternative perspectives
into data or results {[}14, 16, 20{]}. Building this ``geography'' into
visualizations aims for sophisticated tools to subset the actions
available to the user at any single moment, hoping to maintain interface
simplicity {[}54{]}. This tool's summary charts surrounding the
timeseries display seek to demonstrate this potential navigational
method.

\subsubsection{Rumors}\label{rumors}

The use of subtle hints low within visual hierarchy gestures towards
additional optional content and seeks to allow users to ``read at
multiple levels'' without interrupting the regular interaction flow
through the broader tool {[}17, 69, 77{]}. This project's tooltip
markers attempt to provide this unobtrusive prompting.

\subsubsection{Environmental
Explanations}\label{environmental-explanations}

Mirroring the structure of the model into the structure of the tool
affords model mechanics explanations and tries to enable readers to
``play with the premise and assumptions of various claims'' {[}77{]}.
The details tab and, in particular, the assumptions drawers try to
embody this context setting approach.

\subsubsection{Domain Specific Language}\label{domain-specific-language}

Finally, we also leverage a domain specific language to afford agility
and in-situ hot reload, an approach also often considered in games
{[}25, 82, 85{]}.

\subsection{Technical measures for
privacy}\label{technical-measures-for-privacy}

To further this software's stance regarding the avoidance of collecting
information about policy makers and their insights within the sensitive
context of treaty negotiation, the decision support web application
takes a number of privacy-preserving technical steps. Future work in the
policy space should also consider the privacy aspects of not just
research methods but implementation.

\subsubsection{Single page application}\label{single-page-application}

We provide a single page application where computation runs client-side
so that navigation between elements does not generate network requests
that may reveal intentions or activity, a privacy-preserving practice
paired with also avoiding use of technologies like ``beacons'' which
would reveal the questions being explored or results being considered by
a user {[}44{]}.

\subsubsection{Reduced logging}\label{reduced-logging}

We disable or curtail certain web provider defaults for logging, in
particular strictly limiting use of IP addresses to only what is
required for basic operation including traffic routing, security, abuse
prevention, and reliability {[}21{]}.

\subsection{Additional future work}\label{additional-future-work}

Finally, we highlight additional streams of future work.

\subsubsection{Role of AI optimization}\label{role-of-ai-optimization}

This tool could try to use multi-variate optimization to propose
solutions to users, offering its own ``policy answer'' alongside
user-generated suggestions {[}10{]}. However, using the video games
lens, doing so could create ludonarrative dissonance. As an extension of
the project's existing modeling, this hypothetical feature would see the
software claiming to invite outside expertise and exploration (with
computation as a ``helper'' in finding a policy design) while
simultaneously offering a mechanic providing ``optimized'' solutions
(tool as the source of the ``right'' policy design).

In this case of interacting with actual policy negotiation, this project
intentionally excludes this AI-driven auto-optimization option despite
access to the machinery required. If policy makers offer expertise
necessary for finding solutions {[}71{]}, visualization and game design
allow that knowledge to participate in co-creation, potentially reaching
beyond the limits of blind mathematical optimization. Nevertheless,
future work may consider if different design contexts could benefit from
this kind of AI-generated solution or may offer studies on evolving
policy maker perceptions of these algorithmic designs.

\subsubsection{Mediation}\label{mediation}

By affording manipulation of systems and financial assumptions, this
tool answers Victor's call for explorable explanations to allow the
questioning of modeling {[}77{]}. Even so, this dialectic experience
executes within the existing treaty process and does not consider how
interactive software may mediate or alter that process' structure
itself. While tools to actively guide discussion and resolution of
disagreements between stakeholders represent a different goal, this
possibility of mediated disagreement resolution remains a possible area
for future research.

\subsubsection{National and sub-national
usage}\label{national-and-sub-national-usage}

We recognize opportunities for future use of this tool or similarly
designed software at the national or sub-national (such as state) level.
This may operate in a context similar to implementation plans from other
international treaties such as the Kigali Amendment {[}47{]}. However,
these opportunities may become more clear after continued treaty
negotiations.

\subsubsection{Human subjects research}\label{human-subjects-research}

This case study considers quality assurance and improvement activity for
a single tool and these designs. That said, while this simulation
currently remains constrained by the nature of ongoing treaty processes
and a desire to avoid collecting information about individuals within
this sensitive context, later research could seek generalizable
knowledge through human subjects research approaches like MILC {[}64{]}
which may be more appropriate outside active negotiation. For those
efforts, we caution that this interactive tool does not serve the
general public and, instead, focuses on active and often knowledgeable
treaty participants {[}71{]}. Therefore, it remains possible that this
unique target audience may engage with this tool differently than other
users such as general members of the public {[}65{]}. Indeed, situated
techniques often respond to ``domain experts'' with their objectives and
backgrounds mediating experience {[}64{]}. Later research hoping to
build generalizable knowledge should ensure study takes place in-situ
but this study also recognizes a potential opportunity for additional
methods formulation for this kind of active policy negotiation context.

\bigskip 

\section{Conclusion}\label{conclusion}

Where the rigidity of static media limits their mutual embrace, this
case study considers interactivity to facilitate the valuable
conversation between science and policy, a necessary dialogue emphasized
by prior work {[}49, 71{]}. Adding to the early discourse between video
game and information design, we offer a documented open source example
of these ``explorable explanation'' ideas {[}27, 77{]} and how they
influenced a practical tool with real-world use in an international
treaty negotation context. Additionally, we also propose a possible
ludic vocabulary for building these experiences for evidence-based
policy, approaches we found useful in this specific tool's structural
design.

For those facilitating similar machine-mediated conversation, consider a
concept from Jesse Schell central to games: experiences are co-created
by game and player {[}61{]}. In this framing, like games {[}55{]}, this
kind of decision support tool may enable science to be activated and
made more complete by the decision maker when it is brought into the
framework of their policy expertise. This reaches the ultimate ambition
of this tool: using computation to inform policy without prescribing it.

Our global plastics policy simulation tool is available at
\href{https://global-plastics-tool.org}{global-plastics-tool.org}. It is
free and open source. Additionally, a scrolling storytelling experience
for the general public to complement the detailed interface is also
publicly available at
\href{https://plasticstreaty.berkeley.edu/}{https://plasticstreaty.berkeley.edu}.

\bigskip 

\section{Acknowledgements}\label{acknowledgements}

Funding from the Eric and Wendy Schmidt Center for Data Science and
Environment at the University of California Berkeley, the Benioff Ocean
Science Laboratory at the University of California Santa Barbara, the
March Marine Initiative, and the Harris Family Charitable Gift Fund.

\subsection{Authorship}\label{authorship}

\textbf{A. Samuel Pottinger}: conceptualization, methodology, software,
validation, formal analysis, investigation, writing, visualization.
\textbf{Nivedita Biyani}: conceptualization, methodology, software,
validation, formal analysis, data curation, writing. \textbf{Roland
Geyer}: conceptualization, methodology, validation, formal analysis,
investigation, resources, data curation, supervision, funding
acquisition. \textbf{Douglas J McCauley}: conceptualization, validation,
writing, supervision, project administration, funding acquisition,
resources. \textbf{Magali de Bruyn}: writing, software. \textbf{Molly R
Morse}: conceptualization, resources, writing, supervision, project
administration. \textbf{Neil Nathan}: conceptualization, validation,
writing, visualization, resources. \textbf{Kevin Koy}: supervision,
funding acquisition. \textbf{Ciera Martinez}: conceptualization,
validation, investigation, resources, writing, supervision, project
administration.

Reflecting the tool's full credits page at
\url{https://global-plastics-tool.org/\#about-people} at time of
writing, we also acknowledge and thank Elijah Baker for policy research,
Carl Boettiger for early project advice, Kelly Wang for communications,
Linda Nakasone for reporting bugs through open source community process,
and Thought Lab for development of the general public page at
https://plasticstreaty.berkeley.edu/ which complements the tool.
Furthermore, this paper acknowledges and thanks:

\begin{itemize}
\tightlist
\item
  Color Brewer for color schemes {[}13{]}.
\item
  draw.io tools for diagramming {[}2{]}.
\item
  All open source libraries in the project README at
  \url{https://github.com/SchmidtDSE/plastics-prototype}.
\item
  Pandoc for markdown to latex conversion {[}43{]}.
\end{itemize}

Note that, for transparency, the tool also maintains a formal
version-controlled and standards-compliant ``humans.txt''
(\url{https://global-plastics-tool.org/humans.txt}) with comprehensive
and detailed individual credits for the entire project which is current
at time of this paper's submission {[}8{]}. Finally, thank you to policy
makers who have used and considered the tool.

Claude, an AI from Anthropic, was used to help with technical LaTeX
issues and a final proof for the rendered paper but did not
substantively write any of the content which appears in this publication
{[}1{]}.

\subsection{License}\label{license}

Project code is released under the permissive Berkeley Standard
Distribution open source license and invites community contributions
which are also provided under the same BSD terms {[}52{]}.

\subsection{Conflicts of interest}\label{conflicts-of-interest}

We do not have any conflicts to disclose.

\bigskip 

\section*{Works Cited}\label{works-cited}
\addcontentsline{toc}{section}{Works Cited}

\phantomsection\label{refs}

\end{document}